\documentclass[lettersize,journal,10pt]{IEEEtran}
\usepackage{amsmath,amsfonts,mathtools}
\usepackage{pifont}
\usepackage{algorithmic}
\usepackage{algorithm}
\usepackage{array}
\usepackage{textcomp}
\usepackage{stfloats}
\usepackage{url}
\usepackage{verbatim}
\usepackage{graphicx}
\usepackage{gensymb}
\usepackage{cite}
\usepackage{cases}
\usepackage{glossaries}
\usepackage{multirow}
\usepackage{makecell}
\usepackage{color, colortbl}
\definecolor{mixed}{gray}{0.9}
\definecolor{synth}{gray}{0.8}
\usepackage{multicol}
\usepackage[tableposition=top]{caption}
\usepackage{subcaption}
\usepackage{siunitx}
\usepackage[export]{adjustbox}
\usepackage{colortbl}
\usepackage{pgfplots}
\usepackage{pgfplotstable}
\usepgfplotslibrary{fillbetween}
\usepackage{longtable}
\usepackage{xspace}
\usepackage{arydshln}
\usepackage{booktabs}

\usepackage{soul}

\usepackage[font=scriptsize]{subcaption}
\usepackage[font=footnotesize]{caption}

\usepackage[outdir=./]{epstopdf}
\usepackage{tikz}
\pgfplotsset{compat=newest}
\pgfplotsset{plot coordinates/math parser=false}
\newlength\fheight
\newlength\fwidth
\usetikzlibrary{plotmarks, shapes, patterns, decorations.pathreplacing, backgrounds,calc, arrows,arrows.meta,spy, matrix,external}
\usepackage{tikzscale}
\usepackage[utf8]{inputenc}
\usepackage[capitalise]{cleveref}
\crefname{section}{Sec.}{Secs.}
\crefname{figure}{Fig.}{Figs.}

\usepackage{eqparbox}

\DeclareCaptionLabelFormat{andtable}{#1~#2  \&  \tablename~\thetable}

\tikzset{
    export as png/.style={
        external/system call/.add={}{
            && convert -density #1 -transparent white "\image.pdf" "\image.png"
        },
    },
    export as png/.default={200},
}
\newacronym{3gpp}{3GPP}{3rd Generation Partnership Project}
\newacronym{5g}{5G}{5th Generation}
\newacronym{5gc}{5GC}{5G Core}
\newacronym{6g}{6G}{6th Generation}
\newacronym{adc}{ADC}{Analog to Digital Converter}
\newacronym{afbw}{AFBW}{Average Fading Bandwidth}
\newacronym{aimd}{AIMD}{Additive Increase Multiplicative Decrease}
\newacronym{am}{AM}{Acknowledged Mode}
\newacronym{amc}{AMC}{Adaptive Modulation and Coding}
\newacronym{aoa}{AoA}{Angle of Arrival}
\newacronym{aod}{AoD}{Angle of Departure}
\newacronym{ap}{AP}{Access Point}
\newacronym{app}{APP}{Application Layer}
\newacronym{aqm}{AQM}{Active Queue Management}
\newacronym{atis}{ATIS}{Alliance for Telecommunications Industry Solutions}
\newacronym{awgn}{AGWN}{Additive White Gaussian Noise}
\newacronym{balia}{BALIA}{Balanced Link Adaptation}
\newacronym{bdp}{BDP}{Bandwidth-Delay Product}
\newacronym{ber}{BER}{Bit Error Rate}
\newacronym{bler}{BLER}{Block Error Rate}
\newacronym{bf}{BF}{Beamforming}
\newacronym{cad}{CAD}{Computer-Aided Design}
\newacronym{cbr}{CBR}{Constant Bit Rate}
\newacronym{cc}{CC}{Congestion Control}
\newacronym{cdf}{CDF}{Cumulative Distribution Function}
\newacronym{ci}{CI}{Confidence Interval}
\newacronym{cir}{CIR}{Channel Impulse Response}
\newacronym{cn}{CN}{Core Network}
\newacronym{cp}{CP}{Control Plane}
\newacronym{cqi}{CQI}{Channel Quality Information}
\newacronym{crs}{CRS}{Cell Reference Signal}
\newacronym{csirs}{CSI-RS}{Channel State Information - Reference Signal}
\newacronym{d2d}{D2D}{Device-to-Device}
\newacronym{dc}{DC}{Dual Connectivity}
\newacronym{dce}{DCE}{Direct Code Execution}
\newacronym{dci}{DCI}{Downlink Control Information}
\newacronym{dl}{DL}{Downlink}
\newacronym{dmr}{DMR}{Deadline Miss Ratio}
\newacronym{dmrs}{DMRS}{DeModulation Reference Signal}
\newacronym{dod}{DoD}{Department of Defenses}
\newacronym{dray}{D-Ray}{Deterministic Ray}
\newacronym{e2e}{E2E}{End-to-End}
\newacronym{ecn}{ECN}{Explicit Congestion Notification}
\newacronym{ecdf}{ECDF}{Empirical Cumulative Distribution Function}
\newacronym{edf}{EDF}{Earliest Deadline First}
\newacronym{em}{EM}{electromagnetic}
\newacronym{enb}{eNB}{evolved Node Base}
\newacronym{endc}{EN-DC}{E-UTRAN-\gls{nr} \gls{dc}}
\newacronym{epc}{EPC}{Evolved Packet Core}
\newacronym{es}{ES}{Edge Server}
\newacronym{eess}{EESS}{Earth Exploration-Satellite Service}
\newacronym{faa}{FAA}{Federal Aviation Administration}
\newacronym{fcc}{FCC}{Federal Communications Commission}
\newacronym{fdd}{FDD}{Frequency Division Duplexing}
\newacronym{fdma}{FDMA}{Frequency Division Multiple Access}
\newacronym{fr1}{FR-1}{Frequency Range 1}
\newacronym{fr2}{FR-2}{Frequency Range 2}
\newacronym{fr3}{FR-3}{Frequency Range 3}
\newacronym{fray}{F-Ray}{Flashing Ray}
\newacronym{fs}{FS}{Fixed Service}
\newacronym{fss}{FSS}{Fixed Satellite Service}
\newacronym{ftp}{FTP}{File Transfer Protocol}
\newacronym{gmm}{GMM}{Gaussian Mixture Model}
\newacronym{gnb}{gNB}{Next Generation Node Base}
\newacronym{gr}{GR}{Ground Reflection}
\newacronym{harq}{HARQ}{Hybrid Automatic Repeat reQuest}
\newacronym{hetnet}{HetNet}{Heterogeneous Network}
\newacronym{hh}{HH}{Hard Handover}
\newacronym{hol}{HOL}{Head-of-Line}
\newacronym{hpbw}{HPBW}{Half Power Beamwidth}
\newacronym{hqf}{HQF}{Highest-quality-first}
\newacronym{ia}{IA}{Initial Access}
\newacronym{iab}{IAB}{Integrated Access and Backhaul}
\newacronym{ieee}{IEEE}{Institute of Electrical and Electronics Engineers}
\newacronym{imt}{IMT}{International Mobile Telecommunication}
\newacronym{inr}{INR}{Interference to Noise Ratio}
\newacronym{iot}{IoT}{Internet of Things}
\newacronym{is}{IS}{Inter-satellite Service}
\newacronym{itu}{ITU}{International Telecommunication Union}
\newacronym{ked}{KED}{Knife-Edge Diffraction}
\newacronym{kpi}{KPI}{Key Performance Indicator}
\newacronym{ks}{KS}{Kolmogorov–Smirnov}
\newacronym{lcf}{LCF}{Level Crossing Frequency}
\newacronym{lcr}{LCR}{Level Crossing Rate}
\newacronym{los}{LoS}{Line-of-Sight}
\newacronym{lte}{LTE}{Long Term Evolution}
\newacronym{m2m}{M2M}{Machine to Machine}
\newacronym{mac}{MAC}{Medium Access Control}
\newacronym{mc}{MC}{Multi-Connectivity}
\newacronym{mcl}{MCL}{Minimum Coupling Loss}
\newacronym{mcs}{MCS}{Modulation and Coding Scheme}
\newacronym{mec}{MEC}{Mobile Edge Cloud}
\newacronym{metsat}{MetSat}{Meteorological Satellite Service}
\newacronym{mi}{MI}{Mutual Information}
\newacronym{mib}{MIB}{Master Information Block}
\newacronym{mimo}{MIMO}{Multiple Input, Multiple Output}
\newacronym{mlr}{MLR}{Maximum-local-rate}
\newacronym{mls}{MLS}{Microwave Limb Sounder}
\newacronym[plural=\gls{mme}s,firstplural=Mobility Management Entities (MMEs)]{mme}{MME}{Mobility Management Entity}
\newacronym{mmwave}{mmWave}{millimeter wave}
\newacronym{moi}{MoI}{Method of Images}
\newacronym{mpc}{MPC}{Multi Path Component}
\newacronym{mptcp}{MPTCP}{Multipath TCP}
\newacronym{mr}{MR}{Maximum Rate}
\newacronym{mrdc}{MR-DC}{Multi \gls{rat} \gls{dc}}
\newacronym{ms}{MS}{Mobile Service}
\newacronym{mss}{MSS}{Maximum Segment Size}
\newacronym{mtd}{MTD}{Machine-Type Device}
\newacronym{mtu}{MTU}{Maximum Transmission Unit}
\newacronym{nfv}{NFV}{Network Function Virtualization}
\newacronym{nist}{NIST}{National Institute of Standards and Technology}
\newacronym{nlos}{NLoS}{Non-Line-of-Sight}
\newacronym{nr}{NR}{New Radio}
\newacronym{nrmse}{NRMSE}{Normalized Root Mean Square Error}
\newacronym{ns3}{ns-3}{Network Simulator 3}
\newacronym{nsa}{NSA}{Non Stand Alone}
\newacronym{ntia}{NTIA}{National Telecommunications and Information Administration}
\newacronym{ntn}{NTN}{Non Terrestrial Network}
\newacronym{o2i}{O2I}{Outdoor-to-Indoor}
\newacronym{ofdm}{OFDM}{Orthogonal Frequency Division Multiplexing}
\newacronym{osm}{OSM}{OpenStreetMap}
\newacronym{pa}{PA}{Position-aware}
\newacronym{pbch}{PBCH}{Physical Broadcast Channel}
\newacronym{pdcch}{PDCCH}{Physical Downlonk Control Channel}
\newacronym{pdcp}{PDCP}{Packet Data Convergence Protocol}
\newacronym{pdf}{PDF}{Probability Density Function}
\newacronym{pdsch}{PDSCH}{Physical Downlink Shared Channel}
\newacronym{pdu}{PDU}{Packet Data Unit}
\newacronym{per}{PER}{Packet Error Rate}
\newacronym{pf}{PF}{Proportional Fair}
\newacronym{pgw}{PGW}{Packet Gateway}
\newacronym{phy}{PHY}{Physical}
\newacronym{pl}{PL}{Path Loss}
\newacronym{ppp}{PPP}{Poisson Point Process}
\newacronym{prb}{PRB}{Physical Resource Block}
\newacronym{pss}{PSS}{Primary Synchronization Signal}
\newacronym{pucch}{PUCCH}{Physical Uplink Control Channel}
\newacronym{pusch}{PUSCH}{Physical Uplink Shared Channel}
\newacronym{qd}{QD}{Quasi Deterministic}
\newacronym{ra}{RA}{Radio Astronomy Service}
\newacronym{rach}{RACH}{Random Access Channel}
\newacronym{ran}{RAN}{Radio Access Network}
\newacronym[firstplural=Radio Access Technologies (RATs)]{rat}{RAT}{Radio Access Technology}
\newacronym{red}{RED}{Random Early Detection}
\newacronym{rf}{RF}{Radio Frequency}
\newacronym{rfi}{RFI}{Radio Frequency Interference}
\newacronym{rlc}{RLC}{Radio Link Control}
\newacronym{rlf}{RLF}{Radio Link Failure}
\newacronym{rls}{RLS}{Radiolocation Service}
\newacronym{rr}{RR}{Round Robin}
\newacronym{rray}{R-Ray}{Random Ray}
\newacronym{rrc}{RRC}{Radio Resource Control}
\newacronym{rrm}{RRM}{Radio Resource Management}
\newacronym{rs}{RS}{Remote Sensing}
\newacronym{rsrp}{RSRP}{Reference Signal Received Power}
\newacronym{rsrq}{RSRQ}{Reference Signal Received Quality}
\newacronym{rss}{RSS}{Received Signal Strength}
\newacronym{rssi}{RSSI}{Received Signal Strength Indicator}
\newacronym{rt}{RT}{Ray Tracer}
\newacronym{rtt}{RTT}{Round Trip Time}
\newacronym{rw}{RW}{Receive Window}
\newacronym{rx}{RX}{Receiver}
\newacronym{sa}{SA}{standalone}
\newacronym{sack}{SACK}{Selective Acknowledgment}
\newacronym{sap}{SAP}{Service Access Point}
\newacronym{sch}{SCH}{Secondary Cell Handover}
\newacronym{scm}{SCM}{Spatial Channel Model}
\newacronym{scoot}{SCOOT}{Split Cycle Offset Optimization Technique}
\newacronym{sdma}{SDMA}{Spatial Division Multiple Access}
\newacronym{sf}{SF}{Shadow Fading}
\newacronym{si}{SI}{Study Item}
\newacronym{sib}{SIB}{Secondary Information Block}
\newacronym{sinr}{SINR}{Signal-to-Interference-plus-Noise Ratio}
\newacronym{sir}{SIR}{Signal-to-Interference Ratio}
\newacronym{sm}{SM}{Saturation Mode}
\newacronym{snr}{SNR}{Signal-to-Noise Ratio}
\newacronym{son}{SON}{Self-Organizing Network}
\newacronym{sr}{SR}{Space Research Service}
\newacronym{srs}{SRS}{Sounding Reference Signal}
\newacronym{ss}{SS}{Synchronization Signal}
\newacronym{sss}{SSS}{Secondary Synchronization Signal}
\newacronym{sta}{STA}{Station}
\newacronym{subthz}{sub-THz}{sub-TeraHertz}
\newacronym{svd}{SVD}{Singular Value Decomposition}
\newacronym{tb}{TB}{Transport Block}
\newacronym{tcp}{TCP}{Transmission Control Protocol}
\newacronym{udp}{UDP}{User Datagram Protocol}
\newacronym{tdd}{TDD}{Time Division Duplexing}
\newacronym{tdma}{TDMA}{Time Division Multiple Access}
\newacronym{te}{TE}{Transverse Electric}
\newacronym{tfl}{TfL}{Transport for London}
\newacronym{tgad}{TGad}{Task Group ad}
\newacronym{tgay}{TGay}{Task Group ay}
\newacronym{tm}{TM}{Transverse Magnetic}
\newacronym{trp}{TRP}{Transmitter Receiver Pair}
\newacronym{tti}{TTI}{Transmission Time Interval}
\newacronym{ttt}{TTT}{Time-to-Trigger}
\newacronym{tvg}{TVG}{Time Variable Gain}
\newacronym{tx}{TX}{Transmitter}
\newacronym{ue}{UE}{User Equipment}
\newacronym{ul}{UL}{Uplink}
\newacronym{um}{UM}{Unacknowledged Mode}
\newacronym{uma}{UMa}{Urban Macro}
\newacronym{uml}{UML}{Unified Modeling Language}
\newacronym{upa}{UPA}{Uniform Planar Array}
\newacronym{utc}{UTC}{Urban Traffic Control}
\newacronym{vm}{VM}{Virtual Machine}
\newacronym{wbf}{WBF}{Wired Bias Function}
\newacronym{wf}{WF}{Wired-first}
\newacronym{wifi}{Wi-Fi}{Wireless Fidelity}
\newacronym{wigig}{WiGig}{Wireless Gigabit}
\newacronym{wlan}{WLAN}{Wireless Local Area Network}
\newacronym{wrc}{WRC}{World Radiocommunication Conference}
\newacronym{xpr}{XPR}{Cross Polarization Ratio}
\newacronym{sthz}{Sub-THz}{sub-terahertz}
\newacronym{thz}{THz}{terahertz}
\newacronym{aclr}{ACLR}{Adjacent Channel Leakage power Ratio}
\newacronym{fr}{FR}{Frequency Range}
\newacronym{mbs}{MBS}{Multicast and Broadcast Service}
\newacronym{mgws}{MGWS}{Multi-Gigabit Wireless Systems}
\newacronym{uwb}{UWB}{Ultra-Wideband}
\newacronym{mvdss}{MVDDS}{Multi-Channel Video and Data Distribution Service}
\newacronym{dbs}{DBS}{Direct Broadcast Satellite}
\newacronym{cbrs}{CBRS}{Citizen Broadband Radio Service}
\tikzstyle{startstop} = [rectangle, rounded corners, minimum width=2cm, minimum height=0.5cm,text centered, draw=black]
\tikzstyle{io} = [trapezium, trapezium left angle=70, trapezium right angle=110, minimum width=3cm, minimum height=1cm, text centered, draw=black]
\tikzstyle{process} = [rectangle, minimum width=2cm, minimum height=0.5cm, text centered, draw=black, alignb=center]
\tikzstyle{decision} = [ellipse, minimum width=2cm, minimum height=1cm, text centered, draw=black]
\tikzstyle{arrow} = [thick,<->,>=stealth]
\tikzstyle{line} = [thick,>=stealth]
\tikzstyle{darrow} = [thick,<->,>=stealth,dashed]
\tikzstyle{sarrow} = [thick,->,>=stealth]
\tikzstyle{larrow} = [line width=0.1mm,dashdotted,->,>=stealth]

\makeatletter
\def\grd@save@target#1{%
  \def\grd@target{#1}}
\def\grd@save@start#1{%
  \def\grd@start{#1}}
\tikzset{
  grid with coordinates/.style={
    to path={%
      \pgfextra{%
        \edef\grd@@target{(\tikztotarget)}%
        \tikz@scan@one@point\grd@save@target\grd@@target\relax
        \edef\grd@@start{(\tikztostart)}%
        \tikz@scan@one@point\grd@save@start\grd@@start\relax
        \draw[minor help lines] (\tikztostart) grid (\tikztotarget);
        \draw[major help lines] (\tikztostart) grid (\tikztotarget);
        \grd@start
        \pgfmathsetmacro{\grd@xa}{\the\pgf@x/1cm}
        \pgfmathsetmacro{\grd@ya}{\the\pgf@y/1cm}
        \grd@target
        \pgfmathsetmacro{\grd@xb}{\the\pgf@x/1cm}
        \pgfmathsetmacro{\grd@yb}{\the\pgf@y/1cm}
        \pgfmathsetmacro{\grd@xc}{\grd@xa + \pgfkeysvalueof{/tikz/grid with coordinates/major step x}}
        \pgfmathsetmacro{\grd@yc}{\grd@ya + \pgfkeysvalueof{/tikz/grid with coordinates/major step y}}
        \foreach \x in {\grd@xa,\grd@xc,...,\grd@xb}
        \node[anchor=north] at (\x,\grd@ya) {\pgfmathprintnumber{\x}};
        \foreach \y in {\grd@ya,\grd@yc,...,\grd@yb}
        \node[anchor=east] at (\grd@xa,\y) {\pgfmathprintnumber{\y}};
      }
    }
  },
  minor help lines/.style={
    help lines,
    gray,
    line cap =round,
    xstep=\pgfkeysvalueof{/tikz/grid with coordinates/minor step x},
    ystep=\pgfkeysvalueof{/tikz/grid with coordinates/minor step y}
  },
  major help lines/.style={
    help lines,
    line cap =round,
    line width=\pgfkeysvalueof{/tikz/grid with coordinates/major line width},
    xstep=\pgfkeysvalueof{/tikz/grid with coordinates/major step x},
    ystep=\pgfkeysvalueof{/tikz/grid with coordinates/major step y}
  },
  grid with coordinates/.cd,
  minor step x/.initial=.5,
  minor step y/.initial=.2,
  major step x/.initial=1,
  major step y/.initial=1,
  major line width/.initial=1pt,
}
\makeatother

\begin{document}

% \title{Do Reflections Count?\\ A Radio-Frequency Interference Analysis for Spectrum Coexistence at Sub-THz Bands}
\title{Sharing Spectrum and Services in the 7-24 GHz Upper Midband}

\author{Paolo~Testolina,~\IEEEmembership{Member,~IEEE}, Michele~Polese,~\IEEEmembership{Member,~IEEE}, Tommaso~Melodia,~\IEEEmembership{Fellow,~IEEE}
    % <-this % stops a space
    \thanks{P. Testolina, M. Polese, and T. Melodia are with the Institute for the Wireless Internet of Things, Northeastern University, Boston, MA. Email: \{p.testolina, m.polese, melodia\}@northeastern.edu. This work was partially supported by OUSD(R\&E) through Army Research Laboratory Cooperative Agreement Number W911NF-19-2-0221. The views and conclusions contained in this document are those of the authors and should not be interpreted as representing the official policies, either expressed or implied, of the Army Research Laboratory or the U.S. Government. The U.S. Government is authorized to reproduce and distribute reprints for Government purposes notwithstanding any copyright notation herein.}
    % \thanks{Manuscript received xxx, 2022; revised xxx, 2022.}
}

% The paper headers
% \markboth{Journal of \LaTeX\ Class Files,~Vol.~14, No.~8, December~2022}%
% {Shell \MakeLowercase{\textit{et al.}}: A Sample Article Using IEEEtran.cls for IEEE Journals}

% \IEEEpubid{0000--0000/00\$00.00~\copyright~2022 IEEE}

% Remember, if you use this you must call \IEEEpubidadjcol in the second
% column for its text to clear the IEEEpubid mark.

% \makeatletter
% \patchcmd{\@maketitle}
% {\addvspace{0.5\baselineskip}\egroup}
% {\addvspace{-1.5\baselineskip}\egroup}
% {}
% {}
% \makeatother

\maketitle

\begin{abstract}
   The upper midband, spanning 7 to 24 GHz, strikes a good balance between large bandwidths and favorable propagation environments for future \gls{6g} networks. Wireless networks in the upper midband, however, will need to share the spectrum and safely coexist with a variety of incumbents, ranging from radiolocation to fixed satellite services, as well as Earth exploration and sensing. In this paper, we take the first step toward understanding the potential and challenges associated with cellular systems between 7 and 24 GHz. Our focus is on the enabling technologies and policies for coexistence with established incumbents. We consider dynamic spectrum sharing solutions enabled by programmable and adaptive cellular networks, but also the possibility of leveraging the cellular infrastructure for incumbent services. Our comprehensive analysis employs ray tracing and examines real-world urban scenarios to evaluate throughput, coverage tradeoffs, and the potential impact on incumbent services. Our findings highlight the advantages of FR-3 over FR-2 and FR-1 in terms of coverage and bandwidth, respectively. We conclude by discussing a network architecture based on Open RAN, aimed at enabling dynamic spectrum and service sharing.
\end{abstract}

\glsresetall

\begin{tikzpicture}[remember picture,overlay]
  \node[anchor=north,yshift=-10pt] at (current page.north) {\parbox{\dimexpr\textwidth-\fboxsep-\fboxrule\relax}{
  \centering\footnotesize THIS PAPER HAS BEEN SUBMITTED TO IEEE COMMUNICATIONS MAGAZINE. COPYRIGHT MAY CHANGE WITHOUT NOTICE.}};
\end{tikzpicture}

\section{Introduction}
\label{sec:introduction}

% Preamble on 6G and need for spectrum, struggles of FR-2, sub-THz promising for some services but access is far for 6G.

The \gls{6g} of mobile networks aims to achieve higher data rates, lower latency, and support for innovative applications through intelligent and customized networks. To accomplish this, the industry and regulatory bodies are considering new spectrum bands that have been relatively unexplored for networking purposes. \glspl{mmwave}, or \gls{fr} 2, have encountered challenges in the \gls{5g} of 3GPP systems, particularly with coverage and mobility support~\cite{ghoshal2022indepth}. For \gls{6g}, sub-terahertz frequencies are being considered for backhaul extensions, yet these frequencies face similar challenges regarding access and mobility as seen with FR-2.

For these reasons, the wireless community is considering the upper midband ($7-24$ GHz), or FR-3 in 3GPP, as a promising spectrum segment for 6G~\cite{kang2023cellular,cui20236g,38820}. This segment offers substantial bandwidth availability, coupled with a more favorable propagation environment, which simplifies the engineering and management challenges often encountered with FR-2. The primary challenges for mobile systems in FR-3 stem from (i) the diverse requirements across the 17 GHz of bandwidth (e.g., in frontend and antenna design)~\cite{38820}, and (ii) from the need for a certain degree of directionality to improve coverage without harming incumbents~\cite{kang2023terrestrial}.

Furthermore, the spectrum landscape at such frequencies is complex, with sub-bands allocated to a variety of different incumbent services. These include radiolocation, satellite links, broadcasting, and fixed wireless networks~\cite{itu-rr-2020,fcctac2023preliminary,ghosh2023sharing}. In addition, some bands are exclusively allocated for remote sensing and Earth exploration.
Any additional frequency allocation must ensure that the new services operate in a non-interfering basis to existing ones.
Therefore, while the FR-3 band has the potential to offer hundreds of MHz of spectrum for mobile wireless networks, any feasible strategy must take existing incumbents into consideration. This requires sharing resources when possible and avoiding \gls{rfi} when necessary.

In this context, we identify two significant opportunities for spectrum sharing. Firstly, cellular networks are evolving toward becoming more programmable, intelligent, and adaptive systems~\cite{polese2023empowering}. This evolution enables the implementation of dynamic spectrum sharing solutions that not only build upon established strategies, but also allow for spectrum allocations to be updated on a sub-second timescale.  Such advancements are facilitated by the integration of sensing solutions within the network, enabling coordination and visibility across multiple base stations through intelligent controllers and digital twinning, and open interfaces to coordinate with incumbents. 

Additionally, we envision opportunities not only to share spectrum but also to share services, i.e., using the cellular infrastructure for incumbent services such as positioning, navigation, location, and broadcasting. Whenever a cellular network for a new generation of mobile communications is deployed, there is a significant investment to reach or increase base station density usually associated with cellular systems~\cite{polese2023empowering,ghoshal2022indepth}. This means that incumbent services can leverage the widespread presence of cellular access points and pool spectrum with the wireless network, toward improved utilization through statistical multiplexing. 

In this paper, we elaborate on these two strategies and provide an overview of opportunities and challenges around sharing spectrum and services in the upper midband, based on a comprehensive review of the current spectrum landscape and a ray-tracing-based analysis of coverage and \gls{rfi} to incumbents in a urban scenario. We take a decisive first step toward understanding spectrum opportunities for 6G in the upper midband, with an approach that is based on both policy and technology. Our evaluations leverage realistic ray tracing that can be used to digitally twin real-world scenarios. Specifically, compared to prior work~\cite{kang2023cellular,cui20236g}, we consider a real-world deployment with an unprecedented scale and level of details, with up to 50 \glspl{gnb} over a $1.5\times1.5$~km area and 3D building models with sub-meter precision.

Using this platform, we evaluate throughput and coverage tradeoffs across FR-1, FR-2, and FR-3. We also highlight opportunities and challenges around adopting FR-3 for wireless networks. We show that the bandwidth increase makes up for the coverage loss compared to FR-1, and that coverage is about 20\% higher at 12 GHz than in the lowest part of FR-2.
We then review the current spectrum allocation and the coexistence opportunities. For each class of incumbent services, we describe why they need spectrum and whether they could potentially share it, with an evaluation of the possible disruption that \gls{rfi} could introduce to their operations.
For a specific scenario, i.e., a \gls{fs} incumbent in downtown Boston, we numerically evaluate the \gls{rfi} using realistic urban deployment and ray tracing. Results show that the critical interferers can be clearly identified. This suggests that ad-hoc interference management solutions to suppress interference and enable coexistence with the node granularity can play an important role going forward.

Finally, we discuss a network architecture based on Open \gls{ran} that can be used to enable dynamic sharing. These insights provide the foundation for the design and further evaluation of 6G networks in the upper midband.

\begin{figure*}[t]
  \centering
  \setlength\fheight{.55\columnwidth}
  \setlength\fwidth{2\columnwidth}
  \includegraphics[width=2\columnwidth]{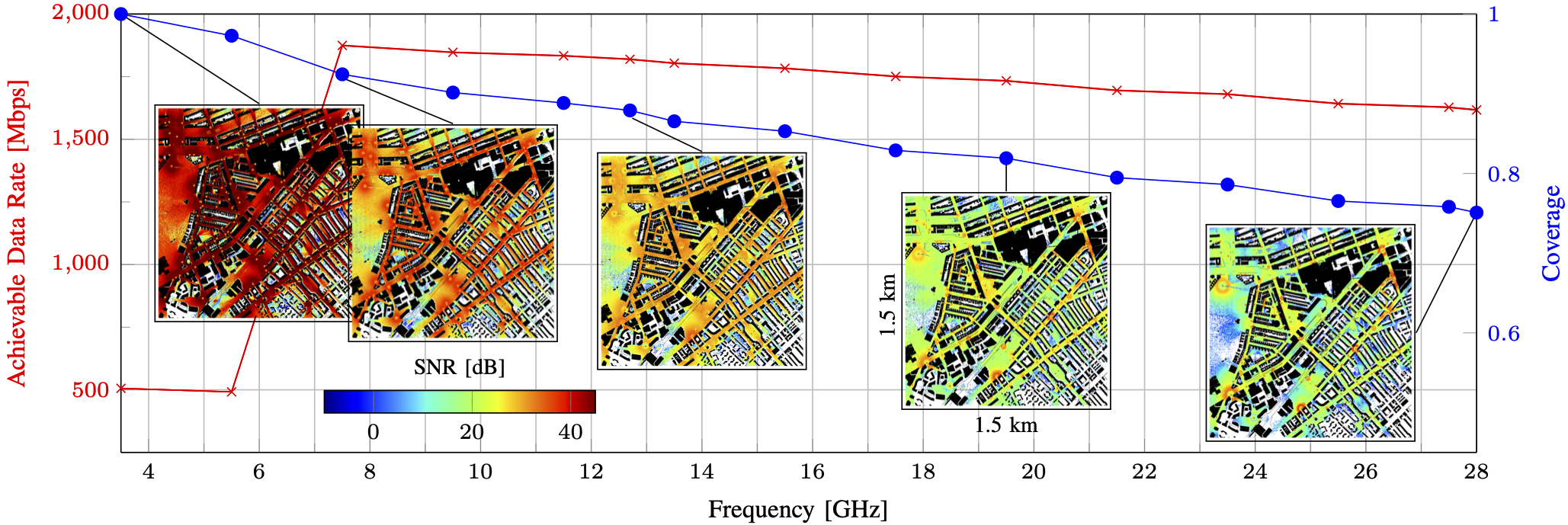}
  \caption{Coverage and throughput for networks deployed at different frequencies in FR-1, FR-2, and FR-3, together with coverage maps for the area considered in the city of Boston. The red dots represent the base station locations.}
  \label{fig:cm_thr}
\end{figure*}

\section{Wireless Cellular Networks in FR-3}
\label{sec:requirements}
% (\myp{I can think/review stuff about this})
% A discussion on the requirements of a system from a network point of view 

This section reviews the challenges and benefits of mobile cellular networks in the upper midband, or FR-3, with insights on the network requirements and propagation characteristics across this spectrum segment. 
FR-3 is indeed extremely diverse: the lower part is used for low-power unlicensed networks (e.g., Wi-Fi 6E) in the U.S., while the upper end is close to what industry traditionally considers mmWave deployments. 
Wireless networks can capitalize on this diversity, enabling high datarate communications (through large swaths of bandwidth, as we discuss in Sec.~\ref{sec:spectrum}) with improved coverage compared to mmWaves.

Figure~\ref{fig:cm_thr} reports coverage maps at different frequencies, generated through ray tracing in an urban scenario in Boston, together with an evaluation of achievable data rates and coverage.
The base stations are deployed in locations currently used to provide cellular service according to open data of the city of Boston, with a density of $\sim18$ base stations per km$^2$. The buildings are modeled with 3D data released by the Boston Planning \& Development Agency. We consider a set of representative carrier frequencies in FR-1, FR-2, FR-3, with a bandwidth of 100 MHz for FR-1 and 400 MHz for FR-2 and FR-3.
We leverage the NVIDIA Sionna~\cite{sionna} ray tracer to simulate the signal propagation, accounting for reflections, as well as diffracted and scattered rays.
The antenna arrays have the same physical aperture and fit as many elements as possible assuming a square \gls{upa} of $\sim40$~mm$^2$ and half lambda separation across elements.
Thus, as the frequency increases, the number of antenna elements increases and the beams are narrower and thus more directional.
The receiver is equipped with a single, omnidirectional antenna.
We consider a beam with $\ang{-12}$ downtilt and steer the array to different directions.

We evaluate the achievable data rate, based on the Shannon capacity equation, and the coverage, computed as the ratio of the area where the \gls{snr} is above 0 dB at the carrier of interest and at 3.5 GHz.
%The bandwidth that can be allocated in different portions of the spectrum plays the most significant role in the average achievable data rate in the area we consider, with an average 3.52$\times$ gain when considering the FR-3 deployments with 400 MHz vs. FR-1 where the allocation is at most 100 MHz.
The bandwidth available at different frequencies plays the most significant role in the average achievable data rate, with an average 3.52$\times$ gain when considering the FR-3 deployments with 400 MHz vs. FR-1 where the allocation is at most 100 MHz.
The main drawback of the higher frequency bands lies in the increasing degradation of the \gls{snr} behind structures that prevent \gls{los} (e.g., bottom left part of each coverage map) and in open spaces further away from the base stations (e.g., the left part of the map), that leads to a slight reduction in coverage.
Nonetheless, Fig.~\ref{fig:cm_thr} shows that existing base stations can provide as much outdoor coverage as a 3.5 GHz deployment in 88\% of the area at 12.7 GHz, dropping at 75\% for 28 GHz. 

Therefore, an FR-3 network (e.g., at 12.7 GHz) can access the same bandwidth of an FR-2 deployment with a more advantageous environment for propagation. This opens up support for high-bandwidth applications with lower costs than at FR-2. Such services include traditional cellular systems with high-capacity hotspots for mobile devices and fixed wireless access, but also enhanced \gls{iab} in static and mobile scenarios. Besides conventional 3GPP solutions, the wide bandwidth can set the cellular network infrastructure up for supporting high-precision positioning and sensing, indoor or outdoor, as well as radionavigation and broadcasting. These applications are already often provisioned with links in the upper midband. A unified air interface that can provide communications and such services introduces opportunities for improved spectrum utilization and multiplexing of spectrum resources.

The main challenges for cellular operations at FR-3 stem from the diversity of devices and propagation at various subbands.
% First, the path loss increases by more than 10~dB when moving from 7 to 24 GHz, as also shown in Fig.~\ref{fig:cm_thr}.
% This reduces the potential coverage when considering the same base station and array densities, or requires more directional links and thus complex beam management procedures.
% At the same time, however, there are no significant absorption patterns that fragment this band, contrary to, for example, sub-terahertz frequencies also considered for 6G.
First, %although there are no significant absorption patterns that fragment this band, contrary to, for example, sub-terahertz frequencies,
the path loss increases by more than 10~dB when moving from 7 to 24 GHz, leading to lower coverage when considering the same base station and array densities.
The challenges extend also to the design of the \gls{rf} frontend itself~\cite{38820}, as providing low and consistent \gls{aclr} across the entire frequency range is non-trivial and can impact coexistence with out-of-band services.
Similarly, antennas and RF frontends will likely require dedicated solutions in different subbands, possibly complicating the \gls{ue} considering the form factor constraints.
Finally, signaling procedures and protocols (from beam management to mobility and scheduling) have to account for the diversity of this spectrum and the need for sharing.

Despite these challenges, there is potential for leveraging FR-3 as an enabler of a diverse set of services from a cellular infrastructure which is easier to engineer and less expensive than a mmWave one. Still, the complex spectrum landscape calls for a careful study and selection of the bands of interest for wireless networking, together with the definition of sharing mechanisms at a spectrum and service level. To develop this, it is paramount to understand who are the current stakeholders (Sec.~\ref{sec:spectrum}) and how they are affected by interference (Sec.~\ref{sec:rfi}).

\section{Current Spectrum Allocation and Coexistence Opportunities}
\label{sec:spectrum}
\begin{figure*}[t]
  \centering
  \setlength\fheight{1.4\columnwidth}
  \setlength\fwidth{1.9\columnwidth}
  \includegraphics[width=\fwidth]{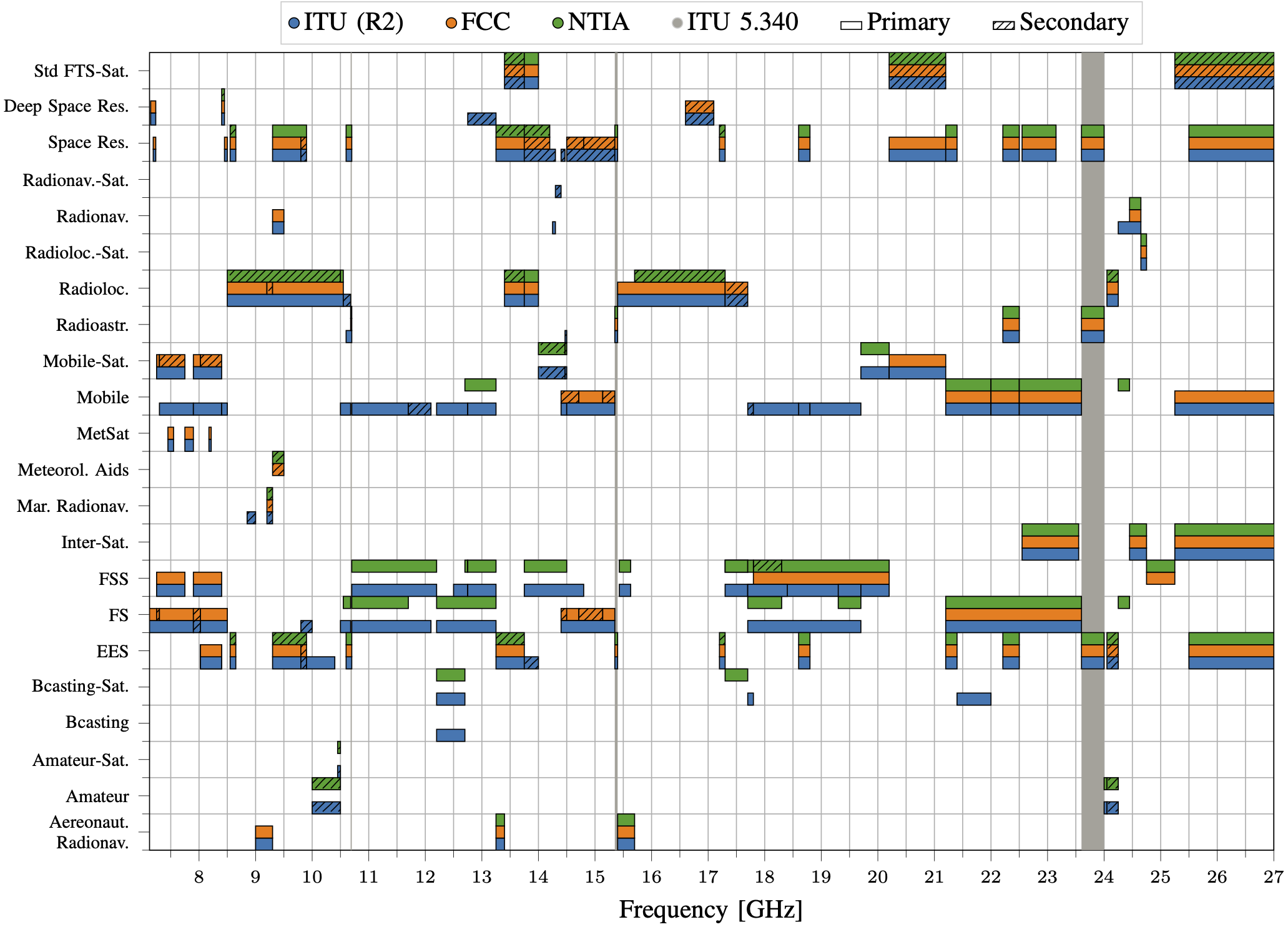}
  \caption{Current spectrum allocations by the \gls{fcc}, the \gls{ntia}, and the \gls{itu} Region 2.}
  \label{fig:regulations}
\end{figure*}

In this section, we present a survey of the current spectrum allocation in the upper midband for Region 2 (the Americas), highlighting coexistence opportunities and challenges.
Figure~\ref{fig:regulations} offers a complete overview of the allocations for Region 2, with further details in~\cite{fcctac2023preliminary}.

\subsection{Non-Scientific Services in \gls{fr}-3}
% \subsubsection{Non-scientific Stakeholders}

  \textbf{Mobile}:
  \gls{ms} already has the largest allocation in \gls{fr}-3, with $11.305$~GHz allocated overall in Region 2, $11.105$~GHz of which according to international regulations and just over 5.822~GHz for federal and non-federal use.
  \cite{itu-r-m2376} investigates the technical feasibility of \gls{imt} in the frequencies between 6 and 100 GHz based on studies carried out by a large number of sector members in different organizations globally, collects measurements and reports different channel models, highlighting that providing \gls{ms} is technically possible.
  % Specifically, it highlights that mobile network prototyping activities are already underway in frequencies such as 11, 15, 28~GHz and that there exists products available for a large variety of bands between 6 GHz and 100 GHz, showing that there is nothing in these bands that inherently prevents their commercial mobile usage.

  \textbf{Fixed Service and Fixed Satellite Service}:
  The point-to-point microwave \gls{fs} occupies the second largest bandwidth in the midband spectrum, with a total of 9.23 GHz reserved by \gls{itu} for primary, distributed in eight chunks that range from 180 MHz to 2.75 GHz of bandwidth.
  With almost 60\% overlap to the \gls{fs}, the \gls{fss} has the third largest bandwidth allocation ($7.40$~GHz) in the \gls{itu} allocations for Region 2.
  Due to its static nature, the coexistence of \glspl{fs} is relatively easy to manage.
  For this reason, there is a substantial overlap with other services, particularly \gls{fss} and Mobile, but also more sensitive ones, e.g., \gls{eess} and Space Research, discussed next.
  However, large chunks of the \gls{fs} frequencies are used by sensitive applications, e.g., by the \gls{faa} to connect remote long-range aeronautical radio-navigation radars to air traffic control centers ($7.125-8.5$~GHz) or by the \gls{dod}.
  % Notably, Resolution 175 of \gls{wrc}-19 ``Use of International Mobile Telecommunications systems for fixed wireless broadband in the frequency bands allocated to the fixed service on a primary basis'' to enable the use of \gls{imt} systems for fixed wireless broadband is under review at \gls{wrc}-23.
  % \begin{itemize}
  %     \item \cite{itu-r-f758} the main \gls{itu} document for the coexistence with \gls{fs}
  %     \item Resolution 175 of \gls{wrc}-19: Use of International Mobile Telecommunications systems for fixed wireless broadband in the frequency bands allocated to the fixed service on a primary basis~\cite{itu2022wrc23agenda}
  % \end{itemize}

  \textbf{Radiolocation}:
  % ITU-R-RS.1281 "Protection of stations in the radiolocation service from emissions from active spaceborne sensors in the band 13.4-13.75 GH"
  In the U.S., most of the \gls{rls} allocation is to federal service and defense radars, e.g., including search, tracking, and missile and gun fire-control radars, some of it shipborne or in selected locations.
  Thus, protection and sharing mechanism with \gls{ms} could be enabled in selected areas and on an opportunistic basis, e.g., prioritizing military services but allowing coexistence as in the \gls{cbrs} band, as these bands are currently under-utilized~\cite{atis2023}.
  $4.750$~GHz are allocated to the \gls{rls} (fourth largest).
  
% \end{itemize}

\subsection{Scientific Stakeholders in FR-3}
The wireless spectrum is a key resource not only for commercial users but also for scientific research.
Among the ``scientific'' services, we include the (Deep) \gls{sr}, the \gls{ra}, the \gls{eess}, and the \gls{metsat}. The \gls{is} is also currently used mainly used by scientific users.
The bands allocated to these services are used both to observe the spectrum at specific frequencies and to distribute the corresponding data and telemetry, telecommand, and ranging signals among ground and spaceborne stations~\cite{itu2017meteohandbook}.

In general, both communication and sensing rely on satellites and ground stations with high sensitivity and precision instruments. Any change in their frequency assignment requires upgrading satellite transceivers and instruments, which is not feasible in the short or medium time frame and costly in the long term.
Furthermore, the frequencies used for the scientific observation are selected due to the presence of specific interactions or phenomena that can be only observed in those bands, or due to their particular characteristics of the band.
Thus, other services in these bands must meet the protection criteria for coexistence with the scientific incumbents, which need to be protected from harmful interference.

 \textbf{Space Research}:
  The \gls{sr} is ``a radiocommunication service in which spacecraft or other objects in space are used for scientific or technological research purposes''~\cite{itu-rr-2020}.
  In \gls{fr}-3, it consists mainly of transmission links between Earth stations, satellites, and spacecrafts.
  The \gls{itu} Regulations for Region 2 allocate 14 bands for a total of 5.635~GHz for primary use.
  
  \textbf{Earth Exploration Satellites, Inter Satellite, and MetSat}:
  Earth exploration (or remote sensing) satellites are used to gather data about the Earth and its natural phenomena.
  \gls{eess} sensors are divided into active, which obtain information by transmitting radio waves and then receiving their reflected energy, and passive sensors, which measure the electromagnetic energy emitted, absorbed, or scattered by the Earth’s surface or atmosphere.
  The ITU Regulations for Region 2 allocate 15 bands for a total of 5.065~GHz for primary use, including the \gls{metsat}, a sub-service of the \gls{eess} dedicated to meteorological purposes~\cite{itu-rr-2020}.
  
  \textbf{Radio Astronomy}:
  The \acrlong{ra} mainly relies on ground radio telescopes, that observe specific frequencies to record events and survey the universe.
  To do that, they employ large antennas and some of the most sensitive instruments on the planet. For this reason, the \gls{itu} also defined some ``zero emission'' zones around \gls{ra} sites under rule 5.340 of the \gls{itu} Radio Regulations \cite{itu-rr-2020}.

\begin{figure*}[t]
  \centering
  \setlength\fheight{.3\columnwidth}
  \setlength\fwidth{2\columnwidth}
  \includegraphics[width=\fwidth]{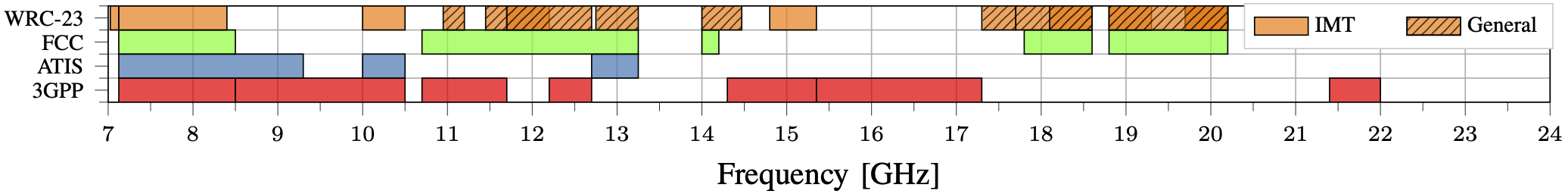}
  \caption{Candidate bands by \gls{fcc}~\cite{fcctac2023preliminary}, ATIS~\cite{atis2023}, and \gls{3gpp}~\cite{38820}, and allocation changes discussed during \gls{wrc} 23 (overlapping regions indicate different resolutions are under consideration for those frequencies).}
  \label{fig:candidate_bands}
\end{figure*}

\subsection{Interference analysis, current coexistence and protection criteria}
Frequency sharing and coexistence between multiple, diverse services is already in place, as shown in \cref{fig:regulations}.
Specifically, \gls{ms} already coexists with several services.
We report here a brief overview of the existing sharing criteria and regulations according to the \gls{itu} Recommendations, highlighting some prominent examples.
Table 5 of \cite{itu2022mobilehandbook} lists the \gls{itu} \gls{imt} frequency spectrum sharing studies for different services, but only for sub-6~GHz frequencies, highlighting the need for further studies and analyses.

Another notable example of sharing is with the communication services among the \gls{eess}, e.g., in the $8.025-8.4$~GHz band, where also \gls{fs}, \gls{fss}, \gls{metsat} and \gls{ms} are allocated. %\cite{itu-r-sa1277}.
Similarly, \gls{eess} bands used for sensing applications have been successfully shared with ground, non-scientific services, although extra requirements are necessary to protect the incumbent (e.g., \cite{itu-r-1803} for $10.6-10.68$~GHz (passive)).

Differently, for the more sensitive \gls{sr} and \gls{ra} services, \gls{los} transmitters are identified as the main potential interference sources.
The main protection against interference for both services is the remote location of the ground stations, and the definition of coordination areas where IMT-2020 can still be deployed after agreement is obtained with the \gls{sr} or \gls{ra} operator.
%~\cite{itu-r-sa2142,itu-r-ra1031}.
As such, ground services are not considered a major threat, unless they employ high-power and high-gain transmitters.
%~\cite{itu-r-sa1016}.
On the contrary, air- and space-borne transmitters are identified as major threats.
%~\cite{itu-r-sa1016,itu-r-sa1155,itu-r-ra1237}.
% \cite{itu-r-sa2142} offers a reference for interference analysis and for the definition criteria of the coordination areas.
% Notably, the aggregate effect from multiple interferers near the earth station is neglected, as the probability of several \glspl{gnb} concurrently pointing towards the incumbent. 
%
% Criteria for the coexistence of \gls{fs} with further services are listed in~\cite{itu-r-f758}.
%
Besides the regulation and recommendations of the specific bands, it is important to consider interference to services in adjacent bands, according to the \gls{aclr} requirements.

Furthermore, besides traditional coexistence solutions, the mobile infrastructure can be leveraged to enable non-mobile services.
An example of this is the \gls{3gpp} \gls{mbs}, where media streaming services can exploit the existing mobile infrastructure with minimal adjustments for on-demand services.
Beyond the broadcast industry, other sectors, notably public safety communications, can exploit the cellular network, facilitating the transmission of data to designated destinations.
% \pt{3GPP MBS 
% % https://www.itu.int/en/ITU-T/Workshops-and-Seminars/202004/Documents/Zeng%20Qingjun.pptx.pdf?csf=1&e=3cIP67
% % https://www.3gpp.org/technologies/broadcast-multicast1
% % https://www.bostonglobe.com/2023/09/10/business/boston-station-5g-tv-broadcasts/
% }
Similarly, the coexistence with other services, like \gls{fs}, can be enabled if the services run on the same (cellular) network, facilitating also their orchestration and coordination, as we discuss in Sec.~\ref{sec:oran}.

\subsection{Candidate Bands}
The relevance of this portion of the spectrum has been increasing over the past years, and several regulatory and commercial entities have expressed their interest.
\cref{fig:candidate_bands} reports the candidate bands identified by members of the \gls{3gpp}~\cite{38820}, by the \gls{fcc}, and by the \gls{atis}.
Furthermore, we reported the frequencies whose allocation was discussed at \gls{wrc} 23.

According to the current allocation (\cref{fig:regulations}), to the interests and the suggestions expressed by major stakeholders and regulators (\cref{fig:candidate_bands}), and taking into account the preliminary analysis presented in~\cref{sec:requirements} and \cref{sec:rfi}, we highlight as most promising bands:
\begin{itemize}

    \item \textbf{$\mathbf{7.125-8.5}$~GHz:} This frequency range has become particularly appealing after the \gls{fcc} decision to open the 6 ($5.925-7.125$) GHz band to unlicensed use, to provide contiguous band usage.
    Existing services include, in decreasing occupancy order, \gls{fs}, \gls{fss}, Satellite \gls{ms}, \gls{eess} (including \gls{metsat}), and (deep) \gls{sr}, according to international and federal regulations.
    A major operator in the US is the Department of Defense, with about 20\% of the \gls{fs} usage.
    Non-federal use includes unlicensed \gls{uwb}, low-power devices, for automotive usage, e.g., radars, personal mobile, and tracking devices.
    The coexistence of ground cellular networks with fixed, spaceborne, and geolocated and geographically-protected incumbents can be managed, whereas the loss of accuracy and reliability of the \gls{uwb} devices due to \gls{rfi} can pose significant challenges.
    
    \item \textbf{$\mathbf{10-10.5}$ GHz:}
    \gls{eess}, \gls{rls}, and the amateur services are the current users of this band. Similarly to the $7.125-8.5$~GHz range, low-power \gls{uwb} are expected to exploit these frequencies.
    Furthermore, the \gls{rls} and the (active) \gls{eess} in this band consist of radars air- and space-borne radars, respectively.
    A major challenge is the presence of passive \gls{eess} in the nearby $10.6-10.7$~GHz band, which requires particular care due to out-of-band emissions.
    
    \item \textbf{$\mathbf{12.2-13.25}$ GHz:}
    Part of this band has been considered by the \gls{fcc}~\cite{ghosh2023sharing}.
    The existing \gls{dbs} and \gls{mvdss} offer interesting sharing opportunities, as the flexible architecture introduced in \cref{sec:oran} allows hosting these services to the future generation of cellular infrastructure.
     
    \item \textbf{$\mathbf{18.8-20.2}$ GHz:}
    While the propagation characteristics of this band are less favorable than the lower ones, hotspots and \gls{wlan} can benefit of the higher bandwidth, comparable to the ones available at the higher \gls{mmwave} frequencies, while maintaining good coverage (above $80\%$), as shown in~\cref{fig:cm_thr}.

\end{itemize}

% \subsection{Candidate Bands}

% \cite{38820} reports as bands of interest for some commercial operators:
% \begin{itemize}
%     \item 7.125 – 8.5 GHz Orange, ETNO, GSMA, ATU
%     \item 8.5 - 10.5 GHz ATU
%     \item 10.7 - 11.7 GHz GSMA, Etisalat
%     \item 12.2 - 12.7 GHz Dish Network
%     \item 14.3 - 15.35 GHz ETNO, GSMA, ATU
%     \item 15.35 - 17.3 GHz ATU
%     \item 21.4 - 22 GHz Orange
% \end{itemize}
% \cite{atis2023} reports as bands of interest:
% \begin{itemize}
%     \item 7.125 - 9.3 GHz
%     \item 10 - 10.5 GHz
%     \item 12.7 - 13.25 GHz
%     \item 25.25 - 27.5 GHz
% \end{itemize}
% The \gls{fcc} suggests limiting to the 7.125 - 15 GHz part of the spectrum, as it offers more desirable propagation characteristics~\cite{fcctac2023preliminary}:
% \begin{itemize}
%     \item 7.125 - 8.5 GHz for sharing with federal fixed, fixed satellite and mobile satellite services
%     \item 10.7 – 13.25 GHz for sharing with non-federal satellite (there is an NPRM on 12.7 – 13.25 GHz and FNPRM in 12.2 – 12.7 GHz)
%     \item 14.0 - 14.2 GHz for sharing with space research
%     \item 17.8 – 18.6 GHz and 18.8 – 20.2 for sharing with federal satellite. Additional analyses need to be done about commercial satellite use of this part of the spectrum.
% \end{itemize}

\section{RFI Evaluation}
\label{sec:rfi}
\begin{figure}
    \centering
    \includegraphics[width=\columnwidth]{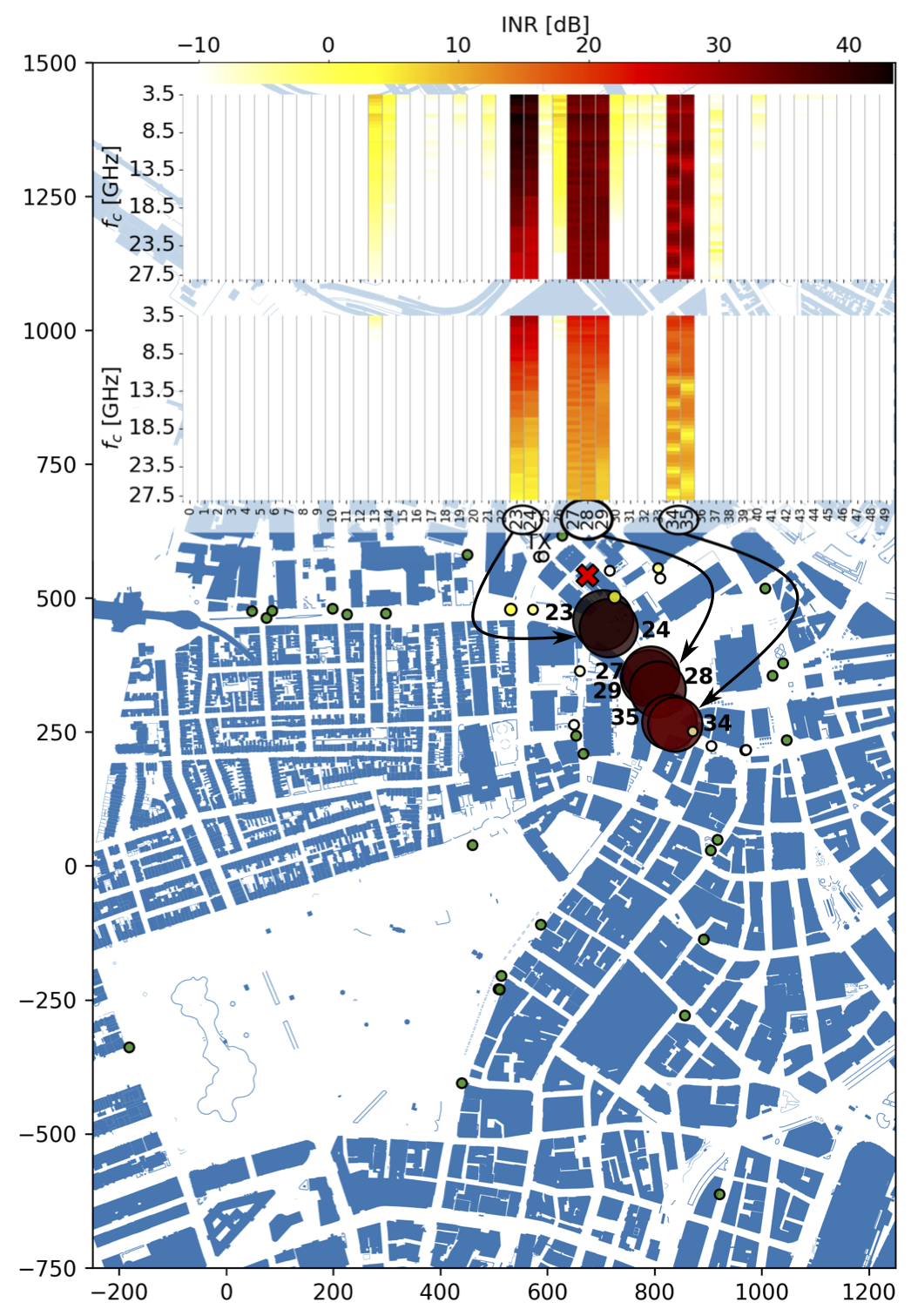}
    \caption{Interference to a \gls{fs} antenna (red cross) from the \glspl{gnb} (dots) in the neighboring area. The heatmaps report the \gls{inr} produced by each \gls{gnb} in the worst (top) and average (bottom) case, over 500 Monte Carlo iterations. The color and size of the markers on the map represent the average \gls{rfi} generated by the \gls{gnb}. The green dots represent the \gls{gnb} whose \gls{inr} does not exceed the $-10$~dB threshold even in the worst case, at any frequency.}
    \label{fig:fs_rfi}
\end{figure}

As mentioned in \cref{sec:introduction}, the frequencies in the broad range of bands under consideration correspond to very different propagation characteristics.
Within the regulation landscape presented in \cref{fig:regulations}, different analyses and sharing opportunities have to be considered, with respect to the existing services and the corresponding frequency, taking into account the specific frequency-service combination.
Thus, we present a \gls{rfi} analysis over the considered spectrum, to assess not only the propagation properties from a communication perspective but also their impact on other services.
In particular, we consider a \gls{fs} incumbent, located on a rooftop in downtown Boston.
We follow the same approach used to generate the coverage maps in \cref{fig:cm_thr}, i.e., the transmitters use a $40$~mm$^2$ square \gls{upa}, while the incumbent employs a single, omnidirectional antenna.

The channel between each interferer and the incumbent is computed using Sionna and accounting for the reflection, diffraction, and scattering of the rays on the buildings and the ground.
We compute the \gls{inr} over a 100~MHz bandwidth at \gls{fr}-1 and 400~MHz bandwidth \gls{fr}-2 and \gls{fr}-3, considering a 9~dB noise figure at the receiver.
\cref{fig:fs_rfi} shows the buildings footprint (light blue) and the \glspl{gnb} (dots) in the considered area.
The \gls{inr} is computed at \gls{fr}-1, \gls{fr}-2, and \gls{fr}-3.
For each frequency, the array of the \glspl{gnb} is steered to a random direction over 500 random Monte Carlo iterations.
The heatmap reports the \gls{inr} over the considered bands in the worst (top) and best (bottom) case, for each \gls{gnb} in the area.
Furthermore, the dot size and color are representative of the average \gls{inr} generated by each \gls{gnb} over the considered spectrum.

As expected, employing lower frequencies and, correspondingly, arrays with fewer antennas, generate higher interference, as the propagation characteristics are more favorable and the communication less directional.
On the contrary, signals at higher frequencies get more attenuated and have a reduced impact on the incumbent, making the case for coexistence.
However, due to directional amplification, the worst-case interference is similar throughout the frequency band, as some \glspl{mpc} to the incumbent might be amplified by the main beam of the \gls{gnb}.

Then, we observe that, out of the 50 \glspl{gnb} in the area, only 7 generate significant interference at the incumbent, as highlighted in the map by the dot color and size.
Notably, the \gls{inr} is not only determined by the distance between the incumbent and the interferer, as there exist \glspl{gnb} closer to the incumbent than, e.g., \gls{gnb} 34 and 35, that generate negligible interference.
Rather, the \gls{inr} is determined by the \gls{los}/\gls{nlos} conditions and by the reflection, diffraction, and scattering of the secondary \glspl{mpc}, that in turn heavily depend on the topology of the environment and on the 3D geometry of the buildings.
This further stresses the need not only for highly accurate channel models but also for detailed and precise digital representations of the propagation environments.
In this context, digital twins can be a useful tool to plan, design, and manage the coexistence of diverse systems in overlapping bands.
For instance, in the case reported in \cref{fig:fs_rfi}, adopting interference suppression techniques at the 7 largest interferers can eliminate almost completely the \gls{rfi} at the incumbent.

% \pt{TODO}
% A preliminary but accurate evaluation of interference (in-band or out-of-band).
% I'd focus on passive (satellite) scientific stakeholders, that are the most vulnerable and have not been considered in the NYU analysis.
% Plot/results proposal:
% \begin{itemize}
% \item interference on fixed services (TV uplinks/downlinks)

% \item satellite? Maybe problems with numerical precision

% \item 
%   % \item numerical evaluation of the probability of blockage of the ground-to-satellite path as a function of the ground node's height, for N (3?) different satellite altitudes and different areas (urban canyon (NYC), urban (Boston), rural (Colorado?)). This might be plots and/or maps that are more appealing, considering the venue $\rightarrow$ \textbf{1 to 3 plots}
%   % \item evaluate the time that a satellite points toward a certain area with its main beam, e.g., Boston (note: requires some work to simulate the satellite/beam trajectory). $\rightarrow$ \textbf{0 to 1 plot}
%   % \item evaluate adjacent band rfi, following the approach proposed by \cite{sormunen2022co}, i.e., simulate the adjacent band as the main one and apply an attenuation according to the adjacent channel interference ratio (ACIR). probably show CDFs as in the paper $\rightarrow$ \textbf{1 plot}
% \end{itemize}

\section{A Network Architecture for Agile Spectrum and Services Sharing in FR-3}
\label{sec:oran}

\begin{figure}
    \centering
    \includegraphics[width=.95\columnwidth]{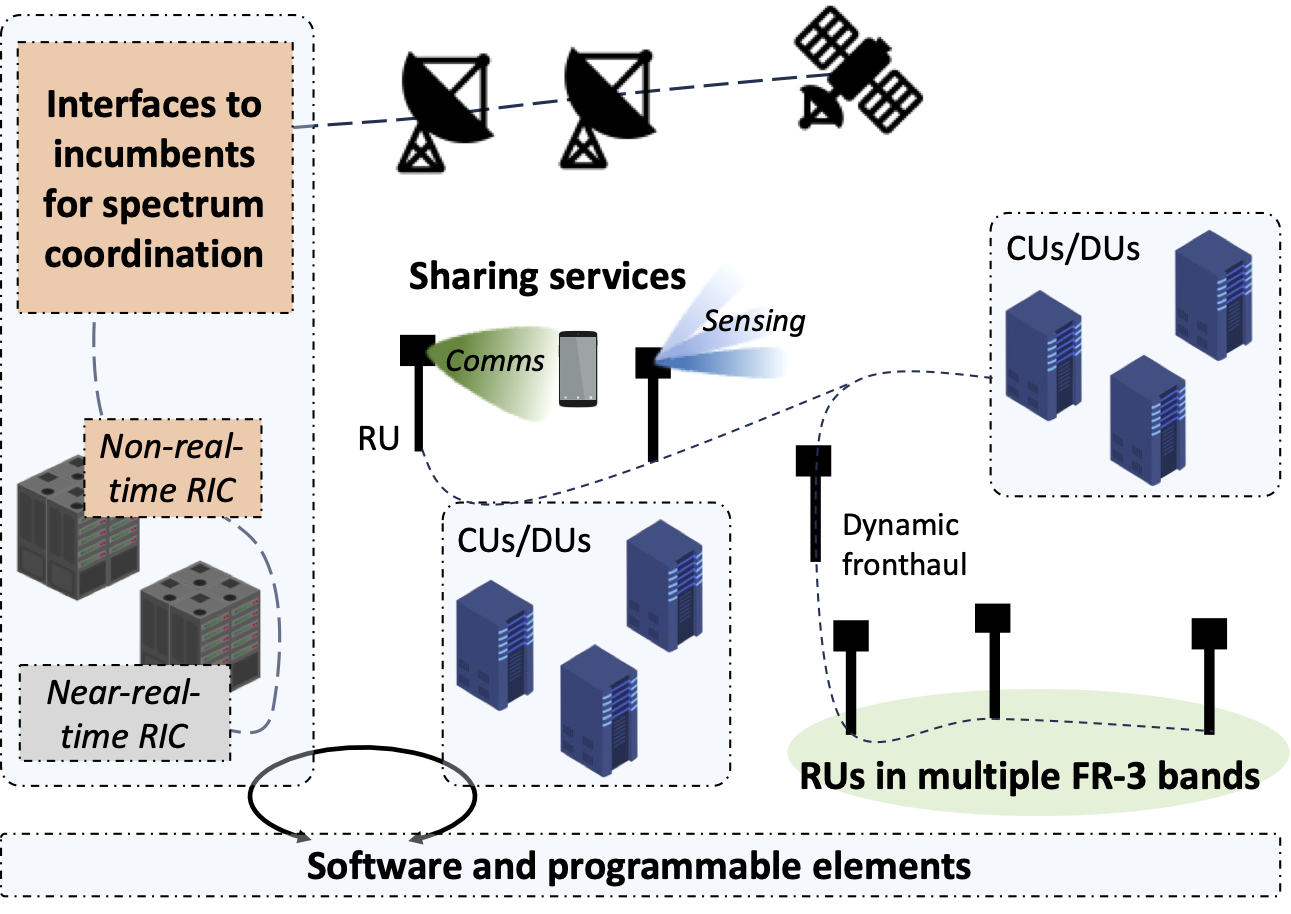}
    \caption{Extension of the Open RAN architecture for FR-3 spectrum and services sharing.}
    \label{fig:oran-for-fr3}
\end{figure}

The analysis above has highlighted the potential of wireless networks in FR-3, together with challenges associated with managing a diverse spectrum landscape and safely integrating and coexisting with current incumbents. This translates into a network architecture which needs to be agile, dynamic, and programmable to efficiently adapt to bespoke conditions in the spectrum utilization and services requested by the users. In this sense, the Open \gls{ran} paradigm, which introduces virtualization, programmability, and plug-and-play disaggregation of the \gls{ran}~\cite{polese2023empowering}, can enable dynamic solutions supporting spectrum and services sharing in FR-3, as we show in Fig.~\ref{fig:oran-for-fr3} and discuss next.

\noindent \textbf{Dynamic Sharing and Coordination with Incumbents.} The Open \gls{ran} architecture comes with programmable, intelligent controllers that serve a dual purpose in this context (left part of Fig.~\ref{fig:oran-for-fr3}). First, they represent an interface between the \gls{ran} and the external world, allowing bi-directional exchange of telemetry and information on the network status but also control and configurations. Second, they allow for a dynamic reconfiguration of the network infrastructure, including a dynamic allocation of the 
% up to the 
spectrum used by the \gls{ran},
% and the fronthaul topology, 
through slicing or dynamic cell configurations. These ingredients allow agile spectrum and infrastructure sharing, with input and/or feedback from current incumbents.

\noindent \textbf{Enabling Dynamic Service Sharing.} Programmability and open interfaces also introduce the possibility to optimize the \gls{ran} and provide bespoke services. As base stations are transitioning toward software, it becomes easier to update their functionality to enable new services using the same infrastructure, spectrum, and potentially waveforms of the cellular network. The services enabled can also be dynamically controlled and adapted over time to match user requests and improve the spectrum utilization, e.g., deploying more spectrum resources for more accurate positioning for safe vehicular operations during rush hour, and scaling back when there is no congestion and the streaming demand for phone calls or video increases.

\noindent \textbf{Manage Diverse RF Requirements.} The Open \gls{ran} flexible and disaggregated network infrastructure can also help addressing the challenges associated with diversity in the RF front-end requirements described in Sec.~\ref{sec:requirements}, at least on the \gls{ran} side. As shown in Fig.~\ref{fig:oran-for-fr3}, the same baseband and protocol stack can indeed be associated with different radio units with various RF characteristics,  including beamforming capabilities and frequency bands, to facilitate sharing and enable new services through the \gls{ran} infrastructure. 

\section{Conclusions}
\label{sec:conclusions}

This paper analyzed spectrum and services sharing between incumbents in the upper midband ($7-24$ GHz) and \gls{6g} networks. We highlighted that FR-3 has a more favorable propagation and potentially similar bandwidth access as FR-2. 
% \gls{fr}-3 presents more favorable propagation conditions compared to the lower mmWave band, thus improving coverage, while potentially providing access to a similar bandwidth and thus data rates. 
% At the same time, 
We discussed how wireless systems will need to share spectrum with existing incumbents, and reviewed coexistence opportunities with different spectrum users, also considering sharing services besides the spectrum itself. We assessed the \gls{rfi} to \gls{fs} in a realistic urban scenario, showing that there exist configurations of the cellular base stations that lead to a very low \gls{inr}, and discussed network architectures for spectrum and services sharing.
Based on our analysis of policy and technology, we conclude that sub-bands $7.125-8.5$, $10-10.5$, $12.2-13.25$, and $18.8-20.2$~GHz are promising candidates for future 6G deployments, because of the coexistence opportunities and the possibility of allocating large chunks of bandwidth for mobile access.

\bibliographystyle{IEEEtran}
\bibliography{bibl}

% Generated by IEEEtran.bst, version: 1.14 (2015/08/26)
\begin{thebibliography}{10}
\providecommand{\url}[1]{#1}
\csname url@samestyle\endcsname
\providecommand{\newblock}{\relax}
\providecommand{\bibinfo}[2]{#2}
\providecommand{\BIBentrySTDinterwordspacing}{\spaceskip=0pt\relax}
\providecommand{\BIBentryALTinterwordstretchfactor}{4}
\providecommand{\BIBentryALTinterwordspacing}{\spaceskip=\fontdimen2\font plus
\BIBentryALTinterwordstretchfactor\fontdimen3\font minus \fontdimen4\font\relax}
\providecommand{\BIBforeignlanguage}[2]{{%
\expandafter\ifx\csname l@#1\endcsname\relax
\typeout{** WARNING: IEEEtran.bst: No hyphenation pattern has been}%
\typeout{** loaded for the language `#1'. Using the pattern for}%
\typeout{** the default language instead.}%
\else
\language=\csname l@#1\endcsname
\fi
#2}}
\providecommand{\BIBdecl}{\relax}
\BIBdecl

\bibitem{ghoshal2022indepth}
M.~Ghoshal, Z.~J. Kong, Q.~Xu, Z.~Lu, S.~Aggarwal, I.~Khan, Y.~Li, Y.~C. Hu, and D.~Koutsonikolas, ``{An In-Depth Study of Uplink Performance of 5G MmWave Networks},'' in \emph{Proceedings of the ACM SIGCOMM Workshop on 5G and Beyond Network Measurements, Modeling, and Use Cases}.\hskip 1em plus 0.5em minus 0.4em\relax Association for Computing Machinery, 2022, p. 29–35.

\bibitem{kang2023cellular}
S.~Kang, M.~Mezzavilla, S.~Rangan, A.~Madanayake, S.~B. Venkatakrishnan, G.~Hellbourg, M.~Ghosh, H.~Rahmani, and A.~Dhananjay, ``{Cellular wireless networks in the upper mid-band},'' \emph{arXiv preprint arXiv:2309.03038}, 2023.

\bibitem{cui20236g}
Z.~Cui, P.~Zhang, and S.~Pollin, ``{6G Wireless Communications in 7-24 GHz Band: Opportunities, Techniques, and Challenges},'' \emph{arXiv preprint arXiv:2310.06425}, 2023.

\bibitem{38820}
3GPP, ``{Study on the 7 to 24 GHz frequency range for NR - Release 16},'' TR 38.820, 2021.

\bibitem{kang2023terrestrial}
S.~Kang, G.~Geraci, M.~Mezzavilla, and S.~Rangan, ``Terrestrial-satellite spectrum sharing in the upper mid-band with interference nulling,'' \emph{arXiv preprint arXiv:2311.12965}, 2023.

\bibitem{itu-rr-2020}
ITU, ``{Radio Regulations},'' ITU-R RR, 2020.

\bibitem{fcctac2023preliminary}
\BIBentryALTinterwordspacing
{FCC Technical Advisory Council}, ``{A Preliminary View of Spectrum Bands in the 7.125 - 24 GHz Range; and a Summary of Spectrum Sharing Frameworks},'' Aug. 2023. [Online]. Available: \url{https://www.fcc.gov/sites/default/files/SpectrumSharingReportforTAC%20%28updated%29.pdf}
\BIBentrySTDinterwordspacing

\bibitem{ghosh2023sharing}
M.~Ghosh, ``{Sharing in the 12 GHz Band},'' \emph{IEEE Wireless Communications}, vol.~30, no.~3, pp. 10--11, 2023.

\bibitem{polese2023empowering}
M.~Polese, M.~Dohler, F.~Dressler, M.~Erol-Kantarci, R.~Jana, R.~Knopp, and T.~Melodia, ``{Empowering the 6G Cellular Architecture with Open RAN},'' \emph{IEEE Journal on Selected Areas in Communications}, pp. 1--1, 2023.

\bibitem{sionna}
J.~Hoydis, S.~Cammerer, F.~{Ait Aoudia}, A.~Vem, N.~Binder, G.~Marcus, and A.~Keller, ``{Sionna: An Open-Source Library for Next-Generation Physical Layer Research},'' \emph{arXiv preprint arXiv:2203.11854}, Mar. 2022.

\bibitem{itu-r-m2376}
ITU, ``{Technical feasibility of IMT in bands above 6 GHz},'' Rec. ITU-R M.2376-0, 2015.

\bibitem{atis2023}
ATIS, ``{Next G Alliance Report: 6G Spectrum Considerations},'' Aug. 2023, white paper.

\bibitem{itu2017meteohandbook}
\BIBentryALTinterwordspacing
ITU and WMO, \emph{Handbook on Use of Radio Spectrum for Meteorology: Weather, Water and Climate Monitoring and Prediction}, 2017. [Online]. Available: \url{https://www.itu.int/dms_pub/itu-r/opb/hdb/R-HDB-45-2017-PDF-E.pdf}
\BIBentrySTDinterwordspacing

\bibitem{itu2022mobilehandbook}
\BIBentryALTinterwordspacing
ITU, \emph{Handbook on International Mobile Telecommunications (IMT)}, 2022. [Online]. Available: \url{https://www.itu.int/dms_pub/itu-r/opb/hdb/R-HDB-62-2022-PDF-E.pdf}
\BIBentrySTDinterwordspacing

\bibitem{itu-r-1803}
------, ``{Technical and operational characteristics for passive sensors in the Earth exploration-satellite (passive) service to facilitate sharing of the 10.6-10.68 GHz and 36-37 GHz bands with the fixed and mobile services},'' Rec. ITU-R RS.1803, 2007.

\end{thebibliography}

\begin{IEEEbiographynophoto}{Paolo Testolina}
[M'23] is a Postdoctoral Researcher at the Institute
for the Wireless Internet of Things, Northeastern University, Boston, since
September 2023. He received his Ph.D. at the Department of Information
Engineering of the University of Padova in 2023. His research interests lie in channel and traffic modeling for wireless communication, and in \gls{rfi} modeling and analysis.

\end{IEEEbiographynophoto}

% \vspace{-1.3cm}
\begin{IEEEbiographynophoto}{Michele Polese}
[M'20] is a Research Assistant Professor at the Institute for the Wireless Internet of Things, Northeastern University, Boston, since October 2023. He received his Ph.D.\ at the Department of Information Engineering of the University of Padova. His research interests are in the analysis and development of protocols and architectures for future generations of cellular networks.
\end{IEEEbiographynophoto}

% \vspace{-1.3cm}
\begin{IEEEbiographynophoto}{Tommaso Melodia}
[F’18] received a Ph.D.\ in Electrical and Computer Engineering from the Georgia Institute of Technology in 2007. He is the William Lincoln Smith Professor at Northeastern University, the Director of the Institute for the Wireless Internet of Things, and the Director of Research for the PAWR Project Office. His research focuses on wireless networked systems.
\end{IEEEbiographynophoto}

\end{document}